\documentstyle[prl,preprint,aps,epsf]{revtex}
\begin{document}

\title{Hidden functional relation in Large-$N$ Quark-Monopole system
at finite temperature}
\preprint{MCTP-01-19}
\author{
D. K. Park\footnote{Email:dkpark@feynman.physics.lsa.umich.edu \\ 
On 
leave from Department of Physics, Kyungnam University, Masan, 631-701, Korea}}
\address{          Michigan Center for Theoretical Physics \\
	 Randall Laboratory, Department of Physics, University of Michigan \\
	 Ann Arbor, MI 48109-1120, USA}

\maketitle

\maketitle
\begin{abstract}
The quark-monopole potential is computed at finite temperature in the context
of $AdS/CFT$ correspondence. It is found that the potential is invariant
under $g \rightarrow 1/g$ and $U_T \rightarrow U_T / g$.  
As in the quark-quark case there exists a maximum
separation between quark and monopole, and $L$-dependence of the potential
exhibits a bifurcation behavior. We find a functional relation
$dE_{QM}^{Reg} / dL = \left[\left(1/E_{(1,0)}^{Reg}(U_0)\right)^2 + 
\left(1/E_{(0,1)}^{Reg}(U_0)\right)^2\right]^{-1/2}$ which is responsible
for the bifurcation. 
The remarkable property of this relation is that it makes a relation 
between physical quantities defined at the $AdS$ boundary through a 
quantity defined at the bulk.
The physical implication of this relation for the
existence of the extra dimension is speculated.
\end{abstract}

\newpage
One of the most important effect of $AdS/CFT$ 
correspondence\cite{mal98-2,ahar00} is that it makes it possible to extract
the highly non-trivial quantum effect in the large N super Yang-Mills 
theories from the classical configuration of string in the $10$-dimensional
$AdS_5 \times S^5$ background. In fact, the expectation value of the rectangular
Wilson loop calculated in the context of $AdS/CFT$ correspondence\cite{mal98-1}
yields an analytical expression for the interquark potential $E_{QQ}$ which
falls off as Coulomb potential, indicating that the theory at the 
$AdS$ boundary is conformally invariant.

Subsequently, Witten\cite{witten98} suggested the generalization of 
$AdS/CFT$ correspondence at finite temperature where temperature is defined 
as an inverse of the compactified Euclidean time. In this context the 
Wilson loop is examined at finite temperature\cite{rey98,brand98}. The 
main difference of the finite temperature case from the zero temperature
one is the appearance of a cusp(or bifurcation point) in the plot of
interquark potential-vs-interquark distance.

Same kind of bifurcation point is also realized in the Euclidean point
particle theory when the finite temperature generalization is considered.
The appearance of the bifurcation point at the point particle theories is
discussed at quantum mechanical\cite{liang98} and field 
theoretical\cite{park99}
levels. In this case the simple functional relation
$dS_E/dP = {\cal E}$, where $S_E$, $P$, and ${\cal E}$ are Euclidean
action, period, and energy of the classical particle, is responsible for
the appearance and disappearance of the bifurcation point. In fact, the 
appearance of the bifurcation point indicates the instability of the 
upper branch in the action-temperature diagram. This means there are 
multiple zero modes at the bifurcation point regardless of the symmetry 
of the underlying theory. This fact is shown explicitly by computing the
spectrum of the fluctuation operator numerically\cite{lee99}. It is 
also possible to prove the multiple zero modes at the bifurcation 
point analytically\cite{min99}. 

Hence, it is natural to ask what kind of functional relation governs the
bifurcation point realized at the Wilson loop calculation in the context
of $AdS/CFT$ correspondence. The answer to this question for the rectangular
Wilson loop is given at Ref.\cite{park01}, where the relation
\begin{equation}
\label{qqcase}
\frac{d E_{QQ}}{dL} = \frac{\sqrt{U_0^4 - U_T^4}}{2 \pi R^2}
\end{equation}
is explicitly derived. Of course, $E_{QQ}$ and $L$ are interquark 
potential and interquark distance, respectively. In Ref.\cite{park01}
the righthand side of Eq.(\ref{qqcase}) is interpreted as a regularized
energy of the trivial string configuration $U=U_0$, {\it i.e.} 
$E^{Reg}(U_0)$. 

It is worthwhile noting that $U_0$ and $U_T$ are locations of minimum point
of string configuration and the horizon respectively. Hence, they are not
defined at the $AdS$ boundary, {\it i.e.} $U = \infty$, while $E_{QQ}$ and 
$L$ are physical quantities defined at the boundary. The remarkable feature
of Eq.(\ref{qqcase}) is that it is a relation between the physical quantities
defined at the $AdS$ boundary via a quantity defined at the bulk space.
This means it is 
impossible to derive Eq.(\ref{qqcase}) if we confine ourselves in the 
4-dimensional world volume without perception of the fifth dimension. 
In this sense the relation (\ref{qqcase}) implies
the physical importance of the extra dimension.
Same kind of the functional relation is derived at the two circular 
Wilson loop case\cite{kim01} and is used for the analysis of the finite
temperature Gross-Ooguri phase transition\cite{gross98,zarem99}. 

In this paper we will show that there is similar functional relation at 
quark-monopole system.
The existence of the same kind of functional relation makes us to expect that
a formula of this type has some universal validity.

The quark-monopole system at zero temperature
is considered in Ref.\cite{mina98} by considering 
3-string junction\cite{schwarz97,gaber97,dasgupta98,sen98,rey97,callan98}.
The quark-monopole-dyon system is also discussed at finite 
temperature\cite{dani98} with highlighting the issues of screening
and clustering. Due to the thermodynamic nature of our hidden functional
relation we believe that it plays more important role when many particles
are involved in the system. In this context it is interesting to 
examine the role of this relation in the multi-particle system located
in the 
external temperature background. We hope to visit this issue in the 
near future.

We start with the classical action of the ($p$,$q$) string worldsheet
\begin{equation}
\label{pqaction1}
S_{(p,q)} = \frac{1}{2 \pi} \int d\tau d\sigma
\sqrt{\left(p^2 + \frac{q^2}{g^2}\right) det G_{MN} \partial_{\alpha}
X^M \partial_{\beta} X^N}
\end{equation}
in the near extremal Euclidean Schwarzschild-$AdS_5\times S^5$ 
background\cite{horo91}
\begin{equation}
\label{ads5}
ds_E^2 = \frac{U^2}{R^2} 
\left[ f(U) dt^2 + dx^i dx^i \right] + 
\frac{R^2 f(U)^{-1}}{U^2} dU^2 + R^2 d\Omega_5^2
\end{equation}
where $f(U) = 1 - U_T^4 / U^4$ and $R = (4\pi g N)^{1/4}$. 
Here, we have chosen $\alpha^{\prime} = 1$
for simplicity. The horizon parameter $U_T$ is propertional to the
external temperature $T$ defined by $T = U_T / (\pi R^2)$\cite{brand98}.

After identifying the world sheet variables as $\tau=t$ and $\sigma=x$, 
one can show easily that for the static case the action $S_{(p,q)}$ becomes
\begin{equation}
\label{pqaction2}
S_{(p,q)} = \frac{\tilde{\tau}}{2 \pi}
\int dx \sqrt{\left(p^2 + \frac{q^2}{g^2}\right)
	      \left[U^{\prime 2} + \frac{U^4 - U_T^4}{R^4}\right]}
\end{equation}
where the prime denotes differentiation with respect to $x$, and 
$\tilde{\tau}$ is the entire Euclidean time interval.

The string configuration we consider is as follows: F-string and D-string
starting from $AdS$ boundary meet with each other at $U=U_0$. For the charge
to be conserved\cite{witten96,stro96} we need (1,1) string starting at
the string junction and ending at the horizon\cite{mal98}. In summary,
the configuration is shown in Fig. 1.

From the world sheet action (\ref{pqaction2}) it is easy to find the 
string configurations of the F-string and D-string:
\begin{equation}
\label{config1}
\frac{U^4 - U_T^4}{\sqrt{U^{\prime 2} + \frac{U^4 - U_T^4}{R^4}}} = 
R^2 \sqrt{U_i^4 - U_T^4}
\hspace{.5cm} i=1,2
\end{equation}
where $i=1(i=2)$ is the result for the F-string(D-string).
As noticed in Ref.\cite{mina98} $U_i$ is not necessarily equal to $U_0$.
Using Eq.(\ref{config1}) it is straightforward to derive
\begin{eqnarray}
\label{distance1}
\Delta L&=& \frac{\alpha R^2 \sqrt{1 - \alpha_T^4}}{U_0}
	\int_{\alpha}^{\infty}
	\frac{dy}{\sqrt{(y^4 - \alpha_T^4)(y^4 - 1)}}
						     \\  \nonumber
L - \Delta L&=& \frac{\beta R^2 \sqrt{1 - \beta_T^4}}{U_0}
	\int_{\beta}^{\infty}
	\frac{dy}{\sqrt{(y^4 - \beta_T^4)(y^4 - 1)}}
\end{eqnarray}
where
\begin{eqnarray}
\label{parameter1}
\alpha&=& \frac{U_0}{U_1} \hspace{1.5cm} \alpha_T = \frac{U_T}{U_1}
							      \\  \nonumber
\beta&=& \frac{U_0}{U_2}  \hspace{1.5cm} \beta_T = \frac{U_T}{U_2}.
\end{eqnarray}

Proceeding as quark-quark case\cite{mal98-1,rey98,brand98,park01} one can 
compute the contribution of F- and D-strings $E_{(1,0)}$ and $E_{(0,1)}$
to the quark-monopole potential:
\begin{eqnarray}
\label{condfNd}
E_{(1,0)}&=&\frac{U_0}{2 \pi \alpha} \int_{\alpha}^{\infty} dy
\sqrt{\frac{y^4 - \alpha_T^4}{y^4 - 1}}     \\   \nonumber
E_{(0,1)}&=&\frac{U_0}{2 \pi g \beta} \int_{\beta}^{\infty} dy
\sqrt{\frac{y^4 - \beta_T^4}{y^4 - 1}}.
\end{eqnarray}
As expected both $E_{(1,0)}$ and $E_{(0,1)}$ are divergent. For the 
regularization we have to substract the quark mass 
$M_q \equiv U_{max} / 2 \pi$ and the monopole mass 
$M_m \equiv U_{max} / 2 \pi g$ from $E_{(1,0)}$ and $E_{(0,1)}$ respectively:
\begin{eqnarray}
\label{regfNd}
E_{(1,0)}^{Reg}&=&\frac{U_0}{2 \pi \alpha}
\left[ \int_{\alpha}^{\infty} dy \left( \sqrt{\frac{y^4 - \alpha_T^4}{y^4 - 1}}
	- 1 \right) - \alpha \right]   \\   \nonumber
E_{(0,1)}^{Reg}&=&\frac{U_0}{2\pi g \beta}
\left[ \int_{\beta}^{\infty} dy \left( \sqrt{\frac{y^4 - \beta_T^4}{y^4 - 1}}
	- 1\right) - \beta \right].
\end{eqnarray}
Hence, combining the contribution of (1,1) string the regularized 
quark-monopole potential $E_{QM}$ becomes the following form:
\begin{equation}
\label{qmpotential}
E_{QM} = E_{(1,0)}^{Reg} + E_{(0,1)}^{Reg} + \frac{U_0 - U_T}{2\pi}
	 \frac{\sqrt{1 + g^2}}{g}.
\end{equation}

The integration in Eq.(\ref{distance1}) and (\ref{regfNd}) are 
analytically carried out in Appendix I and II in terms of the various
elliptic functions. The final result of $L$ and $E_{QM}$ are
\begin{eqnarray}
\label{final}
L&=&\frac{R^2}{U_0} \left[ l(\alpha, \alpha_T) + l(\beta, \beta_T) \right]
						      \\   \nonumber
E_{QM}&=&\frac{U_0}{2\pi} \left[h(\alpha, \alpha_T) + \frac{1}{g}
				h(\beta, \beta_T) \right]
         + \frac{U_0 - U_T}{2\pi} \frac{\sqrt{1 + g^2}}{g}
\end{eqnarray}
where
\begin{eqnarray}
\label{support}
l(x, y)&=&\frac{x}{4} \sqrt{\frac{2(1 - y^2)}{y^2}}
\left[F\left(\Phi(x, y), \kappa(y)\right) - 
      F\left(\Phi(x, y), \kappa^{\prime}(y)\right) \right]
						      \\   \nonumber
h(x, y)&=& - \sqrt{\frac{(x^2 - 1)(x^2 - y^2)}{(x^2 + 1) (x^2 + y^2)}}
						       \\   \nonumber
       &+& \frac{\sqrt{2(1 + y^2)}}{4 x} 
	   \Bigg[(1 - y) F\left(\Phi(x, y), \kappa(y)\right) + 
		 (1 + y) F\left(\Phi(x, y), \kappa^{\prime}(y)\right)
							\\   \nonumber
       & & \hspace{2.5cm}
		 -2 E\left(\Phi(x, y), \kappa(y)\right)
		 -2 E\left(\Phi(x, y), \kappa^{\prime}(y)\right) \Bigg]
							\\   \nonumber
\Phi(x, y)&=&\sin^{-1}\sqrt{\frac{2 x^2 (1 + y^2)}{(x^2 + 1)(x^2 + y^2)}}
							\\   \nonumber
\kappa(y)&=&\frac{1 + y}{\sqrt{2 (1 + y^2)}} \equiv 
\sqrt{1 - \kappa^{\prime 2}(y)}
\end{eqnarray}
and, $F(\phi, k)$ and $E(\phi, k)$ are usual elliptic integral of the
first and second kinds. It is easy to show that $L$ and $E_{QM}$ in
Eq.(\ref{final}) have the correct zero-temperature limit.

Now, let us determine $\alpha$ and $\beta$ from the condition that
the net force at the string junction is zero.
It was conjectured by Schwarz\cite{schwarz97} that such 3-string junctions
with a zero net force corresponds to BPS saturated state and 
subsequently it is verified by world-sheet\cite{dasgupta98} and
space-time\cite{sen98} approaches.

Using string tensions $T^{(1,0)}=\sqrt{U_0^4 - U_T^4} / (2\pi R U_0)$ and
$T^{(0,1)}=\sqrt{U_0^4 - U_T^4} / (2\pi g R U_0)$ one can show directly the 
condition for the zero net force is 
\begin{eqnarray}
\label{BPScon}
& &T^{(1,0)} \left(-\sqrt{\frac{U_1^4 - U_T^4}{U_0^4 - U_T^4}}, 
		  \sqrt{\frac{U_0^4 - U_1^4}{U_0^4 - U_T^4}} \right) +
T^{(0,1)} \left( \sqrt{\frac{U_2^4 - U_T^4}{U_0^4 - U_T^4}},
		 \sqrt{\frac{U_0^4 - U_2^4}{U_0^4 - U_T^4}} \right)
						     \\   \nonumber
& & \hspace{6.0cm}
+ \frac{\sqrt{(1 + g^2) (U_0^4 - U_T^4)}}{2\pi g R U_0} \left(0, -1\right) = 0.
\end{eqnarray}
From Eq.(\ref{BPScon}) one can show $\tan(\theta - \pi/2) = 1/g$ where 
$\theta$ is an angle between D-string and (1,1) string. It is interesting
to realize the fact that the temperature does not affect 
the relative angles between strings.

Solving Eq.(\ref{BPScon}) one can obtain
\begin{eqnarray}
\label{sol12}
\alpha&=&\alpha_0 \equiv \nu \left(\frac{1 + g^2}{\nu^4 + g^2}\right)^{
							      \frac{1}{4}}
							       \\  \nonumber
\beta&=&\beta_0 \equiv \nu \left(\frac{1 + g^2}{1 + g^2 \nu^4}\right)^{
							      \frac{1}{4}}
\end{eqnarray}
where $\nu = U_0 / U_T$. It is worthwhile noting that $\alpha_0 \leftrightarrow
\beta_0$ under the $S$-duality transformation $g \leftrightarrow 1/g$ as
zero temperature case. At zero temperature this is the origin of 
$S$-duality. However, there is a subtle point in $S$-duality at finite
temperature which we will return momentarily.

Using Eq.(\ref{sol12}) the final form of the quark-monopole distance $L$
and potential $E_{QM}$ become
\begin{eqnarray}
\label{final1}
L&=&\frac{R^2}{U_0} \left[l\left(\alpha_0, \frac{\alpha_0}{\nu} \right)
			  + l\left(\beta_0, \frac{\beta_0}{\nu} \right) \right]
						\\   \nonumber
E_{QM}&=&\frac{U_0}{2\pi} \left[h\left(\alpha_0, \frac{\alpha_0}{\nu} \right)
	    + \frac{1}{g} h\left(\beta_0, \frac{\beta_0}{\nu} \right)
	     + \left( 1 - \frac{1}{\nu} \right) \frac{\sqrt{1 + g^2}}{g}
						  \right].
\end{eqnarray}

Now, let us discuss $S$-duality in detail. It is more intuitive to consider 
the zero temperature case first. Taking a zero-temperature 
limit($\nu \rightarrow \infty$) in Eq.(\ref{final1}) yields
\begin{eqnarray}
\label{zerofinal}
L^{(T=0)}&=&\frac{R^2}{U_0} \left[ l_0[(1 + g^2)^{\frac{1}{4}}] + 
				   l_0[\left( \frac{1 + g^2}{g^2} \right)^{
						  \frac{1}{4}} ] \right]
						    \\   \nonumber
E_{QM}^{(T=0)}&=& \frac{U_0}{2\pi}
		  \left[ h_0[(1 + g^2)^{\frac{1}{4}}] + 
			\frac{1}{g} h_0[\left( \frac{1 + g^2}{g^2} \right)^{
						   \frac{1}{4}}]
						   + \frac{\sqrt{1 + g^2}}{g}
							      \right]
\end{eqnarray}
where
\begin{eqnarray}
\label{zerofin}
l_0(x)&=&\frac{x}{\sqrt{2}} \left[ 2 E\left(\sin^{-1}\sqrt{\frac{2}{1 + x^2}}, 
                                            \frac{1}{\sqrt{2}} \right)
                                  - F\left(\sin^{-1} \sqrt{\frac{2}{1 + x^2}},
					    \frac{1}{\sqrt{2}} \right) \right]
                                  - \sqrt{\frac{x^2 - 1}{x^2 + 1}}
							       \\   \nonumber
h_0(x)&=&-\frac{1}{\sqrt{2}x} \left[2 E\left(\sin^{-1}\sqrt{\frac{2}{1 + x^2}},
					     \frac{1}{\sqrt{2}} \right)
                                    - F\left(\sin^{-1}\sqrt{\frac{2}{1 + x^2}},
					     \frac{1}{\sqrt{2}} \right) \right]
                                - \sqrt{\frac{x^2 - 1}{x^2 + 1}}.
\end{eqnarray}
From Eq.(\ref{zerofinal}) one can remove $U_0$ which results in 
\begin{equation}
\label{zerocoul}
E_{QM}^{(T=0)} = \frac{\xi_0}{L^{(T=0)}}
\end{equation}
where
\begin{eqnarray}
\label{zerocoef}
\xi_0&=&\frac{R^2}{2\pi} \left[ l_0[(1 + g^2)^{\frac{1}{4}}]
			       + l_0[\left(\frac{1 + g^2}{g^2} \right)^{
						      \frac{1}{4}} ] \right]
							\\    \nonumber
&\times& \left[ h_0[(1 + g^2)^{\frac{1}{4}}] + \frac{1}{g}
		h_0[\left( \frac{1 + g^2}{g^2} \right)^{\frac{1}{4}} ]
		 + \frac{\sqrt{1 + g^2}}{g} \right].
\end{eqnarray}
Hence, $E_{QM}^{(T=0)}$ falls off as $1 / L^{(T=0)}$, as is required by 
conformal invariance. 
The $R^2 = \sqrt{4 \pi g N}$ proportionality of $\xi_0$ indicates some 
screening of the charge at large distance in zero temperature\cite{mal98-1}.
Furthermore, the coefficient of the Coulomb potential
$\xi_0$ is invariant under the $S$-duality transformation $g \rightarrow 1/g$,
so that $E_{QM}^{(T=0)}$ is also invariant under the same transformation
if and only if $L^{(T=0)}$ is invariant,{\it{i.e.}} $U_0 \rightarrow U_0 / g$
under $g \rightarrow 1/g$. This statement is also easily verified from 
Eq.(\ref{zerofinal}) directly.

Now, let us consider the finite temperature case. Unlike the zero temperature
case it is impossible to remove $U_0$ from Eq.(\ref{final1}) directly. But it
is easy to show that $E_{QM}$ and $L$ are invariant under the transformation
$g \rightarrow 1/g$ and $U_T \rightarrow U_T / g$. It is interesting
that not only the coupling constant but also the temperature parameter are
transformed. This transformation might be the generalized $S$-duality 
transformation at the finite temperature.

From Eq.(\ref{final1}) one can plot the $U_0$-dependence of $L$ which is shown
at Fig. 2. As in the quark-quark case $U_0$-dependence of $L$ 
exhibits monotonic
and non-monotonic behaviors at zero and finite temperature cases respectively.
Fig. 3 shows the $L$-dependence of $E_{QM}$ at various temperature. As in the
quark-quark case there exists an maximum separation $L_{\ast}$ at finite
temperature, which results in the bifurcation. 
Another interesting feature which Fig. 3 indicates is that there are two 
branches of $E_{QM}$ at nonzero temperature, which merge smoothly at 
$L = L_{\ast}$. If the distance between the quark and the monopole is 
greater than $L_{\ast}$, the classical string configuration becomes unstable 
and, as a result, the strings attached to these particles are dropped
on horizon separately. The regularized potential energy of this two 
non-interacting 
particle system is 
\begin{eqnarray*}
E_{iso}^{(Reg)} = - \frac{U_T}{2 \pi} \left(1 + \frac{1}{g} \right).
\end{eqnarray*}
It is interesting to note that $E_{iso}^{(Reg)}$ is equal 
to $E_{QM}$ in Fig. 3 at $L = 0$ in the upper
branch. 
It is easily proved by inserting the $L = 0$ condition $U_T = U_0$
into Eq.(\ref{final1}).
One should note that there exists $L_{\ast \ast}$ at each value of 
nonzero $U_T$ where $E_{QM}$ has same value as that of the isolated system
in the lower branch, which indicates a transition to free particle system.
The interesting points $L_{\ast}$ and $L_{\ast \ast}$ are explicitly 
depicted in Fig. 3 at $U_T = 1.0$.
If the system is given initially at the upper branch of $E_{QM}$ with 
$L < L_{\ast \ast}$, thermal transition should take place to the lower
branch. If, on the other hand, the system is given at either the upper 
or the lower branch with $L_{\ast \ast} < L < L_{\ast}$, the thermal transition
takes place to the two isolated particle system due to its energetical
favor. In fact, this is an exactly same situation with the case of 
Gross-Ooguri phase transition between the catenoid and the two disconnected 
circular Wilson loops\cite{kim01}. Also, same kind of thermal transition 
is discussed in Ref.\cite{dani98} in the more complicated three particle 
system.  

As emphasized
in Ref.\cite{park01,kim01} the non-monotonic behavior of $L$ in Fig. 2 and
the appearance of the cusp in Fig. 3 strongly suggest that there is 
hidden functional relation in the {\bf Y}-junction string system. This hidden
functional relation is explicitly derived at Appendix III, which is 
\begin{equation}
\label{hiddenqm}
\frac{d E_{QM}}{d L} = \frac{1}{2\pi R^2} \sqrt{\frac{U_0^4 - U_T^4}{g^2 + 1}}.
\end{equation}
It is intuitive to compare Eq.(\ref{hiddenqm}) with the functional relation
of the quark-quark case (\ref{qqcase}) explicitly derived at Ref.\cite{park01}:
The right-hand side of Eq.(\ref{qqcase}) is interpreted as the regularized
energy of the trivial string configuration $U = U_0$, {\it{i.e.}}
$E^{Reg}(U_0)$. By the same way the right-hand side of Eq.(\ref{hiddenqm}) can
be expressed in terms of the trivial F- and D-string configurations:
\begin{equation}
\label{interp}
\left[\left(\frac{1}{E_{(1,0)}^{Reg}(U_0)}\right)^2 + 
      \left(\frac{1}{E_{(0,1)}^{Reg}(U_0)}\right)^2 \right]^{-\frac{1}{2}}
\end{equation}
where
\begin{eqnarray}
\label{trivial}
E_{(1,0)}^{Reg}(U_0)&=&\frac{\sqrt{U_0^4 - U_T^4}}{2 \pi R^2}  \\  \nonumber
E_{(0,1)}^{Reg}(U_0)&=&\frac{\sqrt{U_0^4 - U_T^4}}{2 \pi g R^2}.
\end{eqnarray}
Eq.(\ref{interp}) might be a kind of sum rule in the point particle 
anology of the string. 

As in the quark-quark case Eq.(\ref{hiddenqm}) 
represents a functional relation between the physical quantities 
$E_{QM}$ and $L$ defined at the $AdS$ boundary via a quantity which is not
defined at the same boundary. This is the reason why it is impossible to 
derive this kind of relation when we work at four-dimensional world volume.
The functional relations derived at Ref.\cite{park01,kim01} and this paper
may give some insight into the physical importance of the fifth dimension
and yield a conjecture that the fundamental phenomena in our world may be
intricately related to the existance of the extra-dimension.

\vspace{1cm}

{\bf Acknowledgement}: This work was supported by grant No. 
2001-1-11200-001-2 from the Basic Research Program of the Korea
Science and Engineering Foundation.

\newpage
\begin{appendix}{\centerline{\bf Appendix I}}
\setcounter{equation}{0}
\renewcommand{\theequation}{A.\arabic{equation}}
In this appendix we derive $L$ in Eq.(\ref{final}) by integrating 
Eq.(\ref{distance1}) analytically in terms of elliptic functions. Let 
us define $I(\alpha, \alpha_T)$ such that
\begin{equation}
\label{Adef1}
I(\alpha, \alpha_T) \equiv \int_{\alpha}^{\infty} 
\frac{dy}{\sqrt{(y^4 - \alpha_T^4)(y^4 - 1)}}.
\end{equation}
Then Eq.(\ref{distance1}) shows 
\begin{equation}
\label{Aleng}
L = \frac{R^2}{U_0}
    \left[ l(\alpha, \alpha_T) + l(\beta, \beta_T) \right]
\end{equation}
where 
\begin{equation}
\label{Aldefi}
l(x, y) = x \sqrt{1 - y^4} I(x, y).
\end{equation}
Hence, compution of $I(\alpha, \alpha_T)$ completely determines the 
quark-monopole distance.

The first step for the computation of $I(\alpha, \alpha_T)$ is to 
split it into two parts as follows:
\begin{equation}
\label{Asplit}
I(\alpha, \alpha_T) = \frac{1}{\alpha_T^2}
		     \left[I_1(\alpha_T) - I_2(\alpha, \alpha_T) \right]
\end{equation}
where
\begin{eqnarray}
\label{Ai1i2}
I_1(\alpha_T)&=& \int_{\alpha_T}^{\infty}
		\frac{dy}{\sqrt{(1 - y^4) \left(1 - \frac{y^4}{\alpha_T^4}
								    \right)}}
							    \\ \nonumber
I_2(\alpha, \alpha_T)&=& \int_{\alpha_T}^{\alpha}
	       \frac{dy}{\sqrt{(1 - y^4) \left(1 - \frac{y^4}{\alpha_T^4}
								    \right)}}.
\end{eqnarray}
Since $I_1(\alpha_T)$ can be read directly from $I_2(\alpha, \alpha_T)$ by 
taking a $\alpha \rightarrow \infty$ limit, it had better calculate 
$I_2(\alpha, \alpha_T)$ first.

Using formulas $587.00$ and $237.00$ of Ref.\cite{byrd71}, it is 
straightforward to calculate $I_2(\alpha, \alpha_T)$:
\begin{eqnarray}
\label{Aresi2}
I_2(\alpha, \alpha_T)&=& \frac{1}{4} \sqrt{\frac{2 \alpha_T^2}{1 + \alpha_T^2}}
\Bigg[F\left(\sin^{-1} \frac{\sqrt{(\alpha^2 - 1) (\alpha^2 - \alpha_T^2)}}
			    {\alpha^2 - \alpha_T}, 
			    \sqrt{\frac{a + 2}{2 a}}    \right)
								 \\  \nonumber
   & & \hspace{3.0cm} 
    - F\left(\sin^{-1} \frac{\sqrt{(\alpha^2 - 1)(\alpha^2 - \alpha_T^2)}}
			     {\alpha^2 + \alpha_T},
                             \sqrt{\frac{a - 2}{2 a}}   \right) \Bigg]
\end{eqnarray}
where $a = (1 + \alpha_T^2) / \alpha_T \geq 2$.

Taking $\alpha \rightarrow \infty$ limit in Eq.(\ref{Aresi2}) one can obtain
$I_1(\alpha_T)$ as follows:
\begin{equation}
\label{Aresi1}
I_1(\alpha_T) = \frac{1}{4} \sqrt{\frac{2 \alpha_T^2}{1 + \alpha_T^2}}
	       \left[ K\left(\sqrt{\frac{a + 2}{2 a}} \right) - 
		      K\left(\sqrt{\frac{a - 2}{2 a}} \right) \right]
\end{equation}
where $K(\kappa)$ is complete elliptic integral of first kind.

Inserting Eq.(\ref{Aresi2}) and (\ref{Aresi1}) into (\ref{Asplit}) one can
describe $I(\alpha, \alpha_T)$ as a combination of complete and incomplete
elliptic integrals. Using an addition formula
\begin{eqnarray}
\label{Aadd}
F(\theta, \kappa)&+&F(\phi, \kappa) = K(\kappa)  \\   \nonumber
\cot \phi&=&\sqrt{1 - \kappa^2} \tan \theta,
\end{eqnarray}
it is possible to express $I(\alpha, \alpha_T)$ in terms of only 
incomplete elliptic integrals as follows:
\begin{equation}
\label{AfinalI}
I(\alpha, \alpha_T) = \frac{1}{4} \sqrt{\frac{2}{\alpha_T^2 (1 + \alpha_T^2)}}
\left[F(\Phi(\alpha, \alpha_T), \kappa(\alpha_T)) - 
      F(\Phi(\alpha, \alpha_T), \kappa^{\prime}(\alpha_T)) \right]
\end{equation}
where $\Phi(\alpha, \alpha_T)$ and $\kappa(\alpha_T)$ are defined at
Eq.(\ref{support}). Combining Eq.(\ref{Aleng}) (\ref{Aldefi}) and 
(\ref{AfinalI}) it is easy to derive $L$ in Eq.(\ref{final}).
\end{appendix}

\vspace{1cm}
\begin{appendix}{\centerline{\bf Appendix II}}
\setcounter{equation}{0}
\renewcommand{\theequation}{B.\arabic{equation}}

In this appendix we derive $E_{QM}$ in Eq.(\ref{final}) by integrating
Eq.(\ref{regfNd}). Firstly, let us define
\begin{equation}
\label{Bdefi}
J(\Lambda, \alpha, \alpha_T) \equiv 
\int_{\alpha}^{\Lambda} dy \sqrt{\frac{y^4 - \alpha_T^4}{y^4 - 1}}
\end{equation}
where we introduced a cutoff $\Lambda$ which will be taken to infinity at
the final stage of calculation. The first step for the computation of 
$J(\Lambda, \alpha, \alpha_T)$ is also split it into two parts as 
follows:
\begin{equation}
\label{Bsplit}
J(\Lambda, \alpha, \alpha_T) = \alpha_T^2
\left[ J_1(\Lambda, \alpha_T) - J_2(\alpha, \alpha_T) \right]
\end{equation}
where
\begin{eqnarray}
\label{Bj1j2}
J_1(\Lambda, \alpha_T)&=&\int_{\alpha_T}^{\Lambda} dy
\sqrt{\frac{1 - \frac{y^4}{\alpha_T^4}}{1 - y^4}}    \\   \nonumber
J_2(\alpha, \alpha_T)&=& \int_{\alpha_T}^{\alpha} dy
\sqrt{\frac{1 - \frac{y^4}{\alpha_T^4}}{1 - y^4}}.
\end{eqnarray}
As before $J_1(\Lambda, \alpha_T)$ can be read directly from 
$J_2(\alpha, \alpha_T)$ by taking $\alpha \rightarrow \Lambda$ limit. Hence, 
it had better calculate $J_2(\alpha, \alpha_T)$ first.

Next step is to split $J_2(\alpha, \alpha_T)$ into two parts again:
\begin{equation}
\label{Bsplitj2}
J_2(\alpha, \alpha_T) = -I_2(\alpha, \alpha_T) + 
		       \frac{1}{\alpha_T^4} J_3(\alpha, \alpha_T)
\end{equation}
where $I_2(\alpha, \alpha_T)$ is introduced in Eq.(\ref{Ai1i2}) and
\begin{equation}
\label{Bj3}
J_3(\alpha, \alpha_T) = \int_{\alpha_T}^{\alpha} dy
\frac{y^4}{\sqrt{(1 - y^4) \left(1 - \frac{y^4}{\alpha_T^4}\right)}}.
\end{equation}
Since $I_2(\alpha, \alpha_T)$ is calculated explicitly in Appendix I, let
us calculate $J_3(\alpha, \alpha_T)$ here. 

Using formulas $587.03$ and $237.00$ of Ref.\cite{byrd71} it is straightforward
to calculate $J_3(\alpha, \alpha_T)$ after tedious procedure:
\begin{eqnarray}
\label{Bresj3}
& &J_3(\alpha_, \alpha_T)= \frac{\alpha_T^2 \sqrt{\alpha_T}}{4}
\Bigg[ \sqrt{\frac{2}{a}} (a - 1) 
       F\left( \sin^{-1} \frac{\sqrt{(\alpha^2 - 1)(\alpha^2 - \alpha_T^2)}}
			      {\alpha^2 - \alpha_T},
              \sqrt{\frac{a + 2}{2 a}} \right)
						    \\  \nonumber
& &    \hspace{3.0cm}  
+ \sqrt{\frac{2}{a}} (a + 1)
      F\left(\sin^{-1} \frac{\sqrt{(\alpha^2 - 1)(\alpha^2 - \alpha_T^2)}}
			    {\alpha^2 + \alpha_T},
             \sqrt{\frac{a - 2}{2 a}}  \right)  
						    \\   \nonumber
& &    \hspace{3.0cm}
-2 \sqrt{2 a} E\left(\sin^{-1} 
	   \frac{\sqrt{(\alpha^2 - 1)(\alpha^2 - \alpha_T^2)}}
		{\alpha^2 - \alpha_T},
		\sqrt{\frac{a + 2}{2 a}} \right)
						    \\   \nonumber
& &   \hspace{3.0cm}
-2 \sqrt{2 a} E\left(\sin^{-1} 
	   \frac{\sqrt{(\alpha^2 - 1) (\alpha^2 - \alpha_T^2)}}
		{\alpha^2 + \alpha_T},
		\sqrt{\frac{a - 2}{2 a}}  \right)
						    \\   \nonumber
& &    \hspace{3.0cm}
+ \frac{2}{\sqrt{\alpha_T}} 
	   \frac{\sqrt{(\alpha^4 - 1) (\alpha^4 - \alpha_T^4)}}
		{\alpha (\alpha^2 - \alpha_T)}
+ \frac{2}{\sqrt{\alpha_T}}
	   \frac{\sqrt{(\alpha^4 - 1)(\alpha^4 - \alpha_T^4)}}
		{\alpha (\alpha^2 + \alpha_T)}       \Bigg]
\end{eqnarray}
where $a$ is introduced in Appendix I. Inserting Eq.(\ref{Aresi2}) and
(\ref{Bresj3}) into (\ref{Bsplitj2}) it is possible to compute 
$J_2(\alpha, \alpha_T)$ which is not described here due to its lengthy
expression. 

Taking a $\alpha \rightarrow \Lambda$ limit, it is also straightforward to
compute $J_1(\Lambda, \alpha_T)$ which consists of some contribution of 
complete elliptic integrals and $\Lambda$-dependent 
term $\Lambda / \alpha_T^2$. Of course $\Lambda$-dependent term will
be removed 
by regularization. Also using the addition formula (\ref{Aadd}) appropriately,
one obtains the following expression finally:
\begin{eqnarray}
\label{BfinalJ}
J(\Lambda, \alpha, \alpha_T)&=& \Lambda - \alpha 
			\sqrt{\frac{(\alpha^2 -1)(\alpha^2 - \alpha_T^2)}
				   {(\alpha^2+1)(\alpha^2 + \alpha_T^2)}}
							    \\   \nonumber
&+&\frac{\sqrt{2(1 + \alpha_T^2)}}{4}
\Bigg[(1 - \alpha_T) F(\Phi(\alpha, \alpha_T), \kappa(\alpha_T)) +
      (1 + \alpha_T) F(\Phi(\alpha, \alpha_T), \kappa^{\prime}(\alpha_T))
							    \\  \nonumber
& & \hspace{3.0cm}
      -2 E(\Phi(\alpha, \alpha_T), \kappa(\alpha_T)) -
       2 E(\Phi(\alpha, \alpha_T), \kappa^{\prime}(\alpha_T))   \Bigg]
\end{eqnarray}
where $\Phi(\alpha, \alpha_T)$ and $\kappa(\alpha_T)$ are defined at 
Eq.(\ref{support}). From Eq.(\ref{regfNd}), (\ref{Bresj3}) 
and (\ref{BfinalJ}) it is easy
to derive $E_{QM}$ in Eq.(\ref{final}).
\end{appendix}

\vspace{1cm}

\begin{appendix}{\centerline{\bf Appendix III}}
\setcounter{equation}{0}
\renewcommand{\theequation}{C.\arabic{equation}}
In this appendix we derive the functional relation Eq.(\ref{hiddenqm}).
As noticed in Ref.\cite{park01,kim01}, the functional relation is a
thermodynamical anology. Since the thermodynamical relations are usually
realized on the level of first derivative, we compute $dL / d\nu$ and
$dE_{QM}^{Reg} / d\nu$ whose explicit form can be derived using the various 
derivative formulas of the elliptic integrals:
\begin{eqnarray}
\label{C11}
\frac{\partial L}{\partial \nu}&=&
\frac{\sqrt{2} R^2}{4 \nu U_T} \Xi_L
						       \\   \nonumber
\frac{\partial E_{QM}^{Reg}}{\partial \nu}&=&
\frac{U_T}{2\pi} \Xi_E
\end{eqnarray}
where
\begin{eqnarray}
\label{Ccoeff}
\Xi_L&=& -2 \sqrt{\frac{2(\nu^2 + \alpha_0^2)}
		       {(\alpha_0^2 + 1)(\nu^2 + 1)(\nu^2 - \alpha_0^2)}}
       \left[ \alpha_0 \alpha_0^{\prime} \sqrt{\frac{nu^2 - 1}{\alpha_0^2 - 1}}
	      - \nu \sqrt{\frac{\alpha_0^2 - 1}{\nu^2 - 1}}      \right]
							     \\   \nonumber
&-& \frac{\nu \alpha_0^{\prime} - \alpha_0}{\sqrt{\nu^2 - \alpha_0^2}}
  \Bigg[ 2 E\left( \Phi\left( \alpha_0, \frac{\alpha_0}{\nu} \right), 
		   \kappa\left(\frac{\alpha_0}{\nu}\right)  \right)
        +2 E\left(\Phi \left(\alpha_0, \frac{\alpha_0}{\nu} \right),
		  \kappa^{\prime}\left(\frac{\alpha_0}{\nu}\right)  \right)
						      \\   \nonumber
& & \hspace{2.0cm} -
\left( 1 - \frac{\alpha_0(\nu^2 - \alpha_0^2)}{\nu (\nu^2 + \alpha_0^2)} 
								    \right)
F\left(\Phi\left(\alpha_0, \frac{\alpha_0}{\nu} \right), 
       \kappa \left(\frac{\alpha_0}{\nu} \right)         \right)
						       \\   \nonumber
& &   - 
\left(1 + \frac{\alpha_0 (\nu^2 - \alpha_0^2)}{\nu (\nu^2 + \alpha_0^2)}
								   \right)
F\left(\Phi\left(\alpha_0, \frac{\alpha_0}{\nu}\right),
       \kappa^{\prime}\left(\frac{\alpha_0}{\nu}\right)  \right) \Bigg]
+ (\alpha_0 \rightarrow \beta_0, 
     \alpha_0^{\prime} \rightarrow \beta_0^{\prime})
						       \\   \nonumber
\Xi_E&=& \frac{\sqrt{1 + g^2}}{g}
						       \\    \nonumber
     &+& \Bigg[ (\nu \alpha_0^{\prime} - \alpha_0)
	       \bigg[\frac{\sqrt{2(\nu^2 + \alpha_0^2)}}{2 \alpha_0^2 \nu}
		     \left[ E\left(\Phi\left(\alpha_0, \frac{\alpha_0}{\nu}
							     \right), 
                                   \kappa\left(\frac{\alpha_0}{\nu}\right)
							    \right) + 
                            E\left(\Phi\left(\alpha_0, \frac{\alpha_0}{\nu}
							     \right),
                              \kappa^{\prime}\left(\frac{\alpha_0}{\nu}\right)
							     \right)
							     \right]
								 \\   \nonumber
& &    \hspace{2.0cm}
-\frac{\sqrt{2}}{4}
   \frac{\nu^3 - \alpha_0 \nu^2 + \alpha_0^2 \nu + \alpha_0^3}
	{\alpha_0^2 \nu^2 \sqrt{\nu^2 + \alpha_0^2}}
   F\left(\Phi\left(\alpha_0, \frac{\alpha_0}{\nu}\right),
	  \kappa\left(\frac{\alpha_0}{\nu}\right) \right)
							   \\  \nonumber
& &  \hspace{2.0cm}
- \frac{\sqrt{2}}{4}
  \frac{\nu^3 + \alpha_0 \nu^2 + \alpha_0^2 \nu - \alpha_0^3}
       {\alpha_0^2 \nu^2 \sqrt{\nu^2 + \alpha_0^2}}
  F\left(\Phi\left(\alpha_0, \frac{\alpha_0}{\nu}\right), 
	 \kappa^{\prime}\left(\frac{\alpha_0}{\nu}\right) \right)
							  \bigg]
                                                            \\   \nonumber
& &   \hspace{1.0cm}
- \frac{(\nu^2 - 1)(\nu^2 + \alpha_0^2)}
       {\alpha_0 \nu \sqrt{(\alpha_0^4 - 1)(\nu^4 - 1)}} \alpha_0^{\prime}
- \sqrt{\frac{(\alpha_0^2 - 1)(\nu^2 - 1)}{(\alpha_0^2 + 1)(\nu^2 + 1)}}
+\frac{1}{g} \left[ \alpha_0 \rightarrow \beta_0, 
		      \alpha_0^{\prime} \rightarrow \beta_0^{\prime} 
								  \right]
								   \Bigg]
\end{eqnarray}
In Eq.(\ref{Ccoeff}) prime denotes a differentiation with respect to 
$\nu$. Although not immediately obvious from the expression, $\Xi_L$ is
amazingly propertional to $\Xi_E$:
\begin{equation}
\label{Cfinal}
\frac{\Xi_E}{\Xi_L} = \frac{\sqrt{2}}{4 \nu}
\sqrt{\frac{\nu^4 - 1}{1 + g^2}}.
\end{equation}
In fact, Eq.(\ref{Cfinal}) can be proved by comparing all coefficients of 
$\Xi_L$ and $\Xi_E$ one by one. From Eq.(\ref{Cfinal}) it is easy to verify 
Eq.(\ref{hiddenqm}).
\end{appendix}

\begin{figure}
\caption{String configuration we considered in this paper. The (1.1)
string is necessary to conserve (p,q) charge. Since it is known that 
string cannot peneatrate into the horizon, it is attached to $D3$-brane
at $U = U_T$.}
\end{figure}
\vspace{0.4cm}
\begin{figure}
\caption{$U_0$--dependence of $L$ exhibiting the monotonic and 
non-monotonic behaviors at zero and finite temperature.}
\end{figure}
\vspace{0.4cm}
\begin{figure}
\caption{The $L$--dependence of $E_{QM}^{Reg}$ 
exhibiting an bifurcated behavior at finite temperature. This behavior strongly
suggests that there is hidden relation between physical quantities which
is explicitly derived in Eq.(\ref{hiddenqm}).}
\end{figure}

\newpage
\epsfysize=20cm \epsfbox{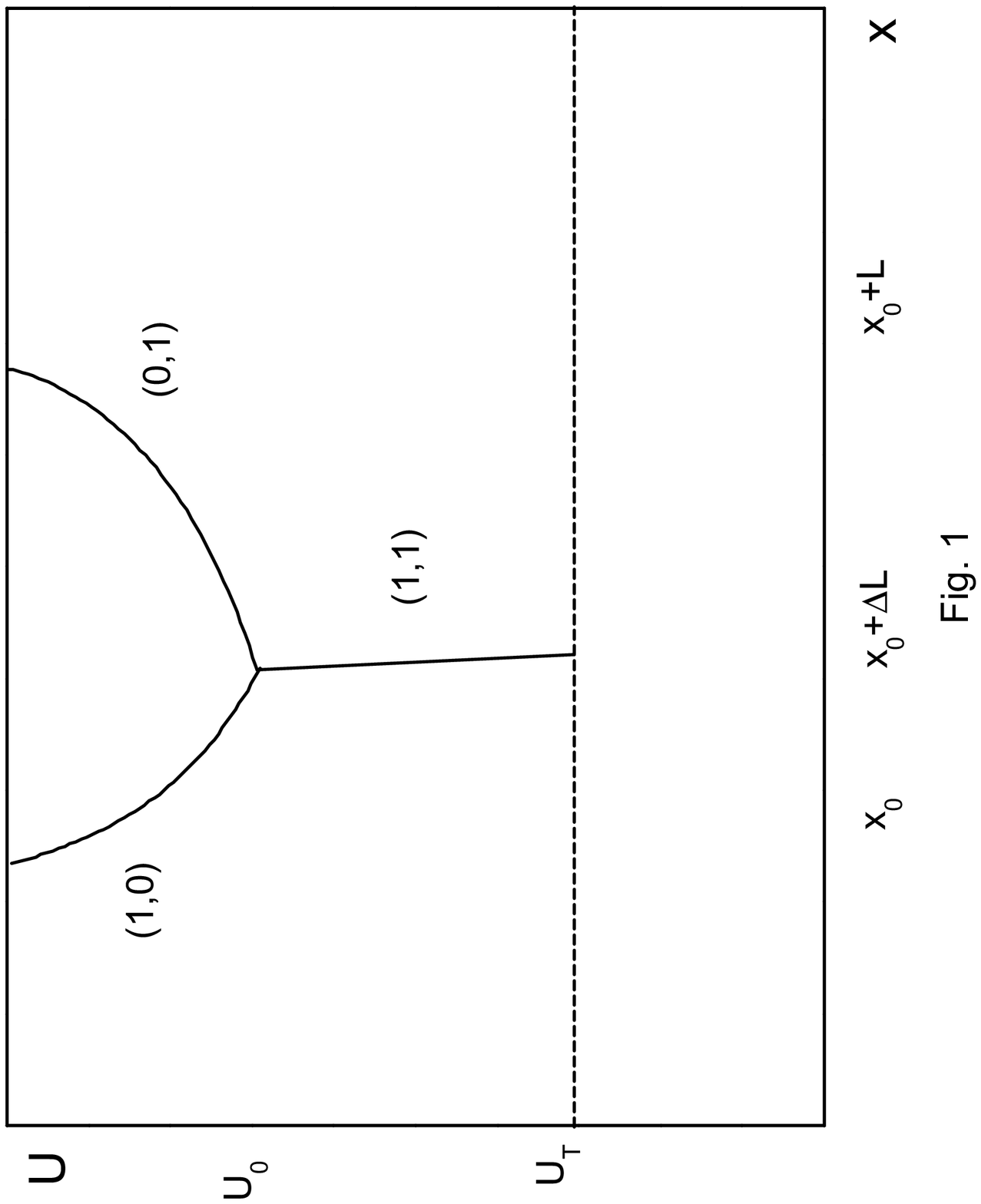}
\newpage
\epsfysize=20cm \epsfbox{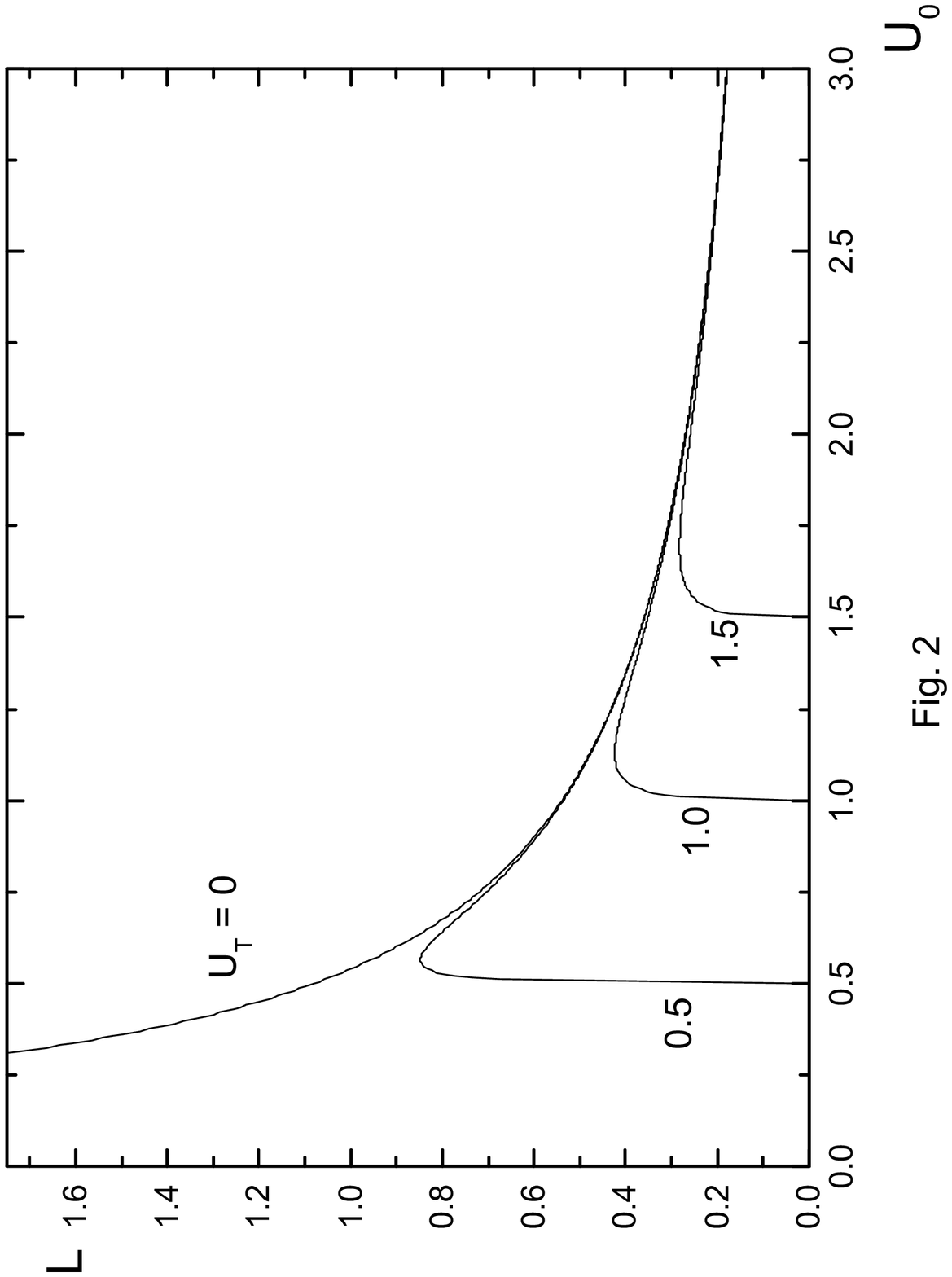}
\newpage
\epsfysize=20cm \epsfbox{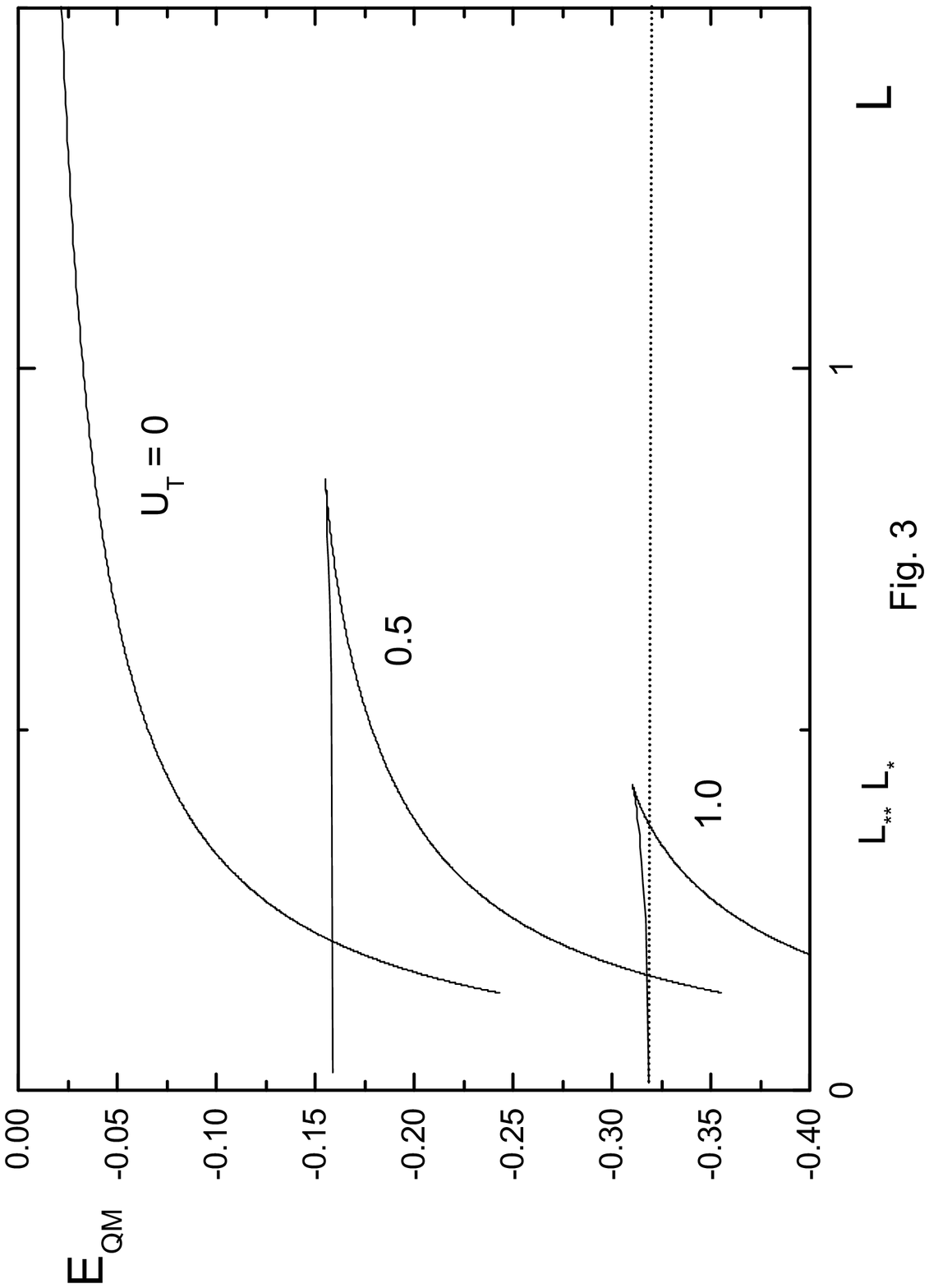}
\end{document}